\documentclass{epl}

\title{Resonating Bipolarons}
\author{J. Ranninger\inst{1} \and A. Romano\inst{2} }

\institute{
  \inst{1} Centre de
Recherches sur les Tr\`es Basses Temp\'eratures, Laboratoire
Associ\'e \`a l'Universit\'e Joseph Fourier,
\\ Centre National de la
Recherche Scientifique, BP 166, 38042, Grenoble
C\'edex 9, France\\
  \inst{2} Dipartimento di Fisica "E.R. Caianiello",
Universit\`a di Salerno, I-84081 Baronissi (Salerno), Italy --
Laboratorio Regionale SuperMat, C.N.R.-I.N.F.M., Salerno (Italy) }

\begin{document}

\maketitle

\begin{abstract}
Electrons coupled to local lattice deformations end up in
selftrapped localized molecular states involving their binding
into bipolarons when the coupling is stronger than a certain
critical value. Below that value they exist as essentially
itinerant electrons. We propose that the abrupt crossover between
the two regimes can be described by resonant pairing similar to
the Feshbach resonance in binary atomic collision processes. Given
the intrinsically local nature of the exchange of pairs of
itinerant electrons and localized bipolarons, we demonstrate the
occurrence of such a resonance on a finite-size cluster made out
of metallic atoms surrounding a polaronic ligand center.

\end{abstract}

The phenomenon of resonant pairing between nucleons, atoms and molecules,
characterized by a strongly enhanced scattering amplitude,  has
received considerable attention in the past few years, primarily in
connection with the study of ultracold dilute atomic gases exhibiting
various forms of condensed macroscopic coherent quantum states. The basic
requirement for such so called  {\it Feshbach
resonant pairing}\cite{Feshbach-58,Timmermans-99} is that two incident
particles have nearly the same energy as that required to form a shallow
bound state between them. The efficiency to convert two uncorrelated atoms
into a molecular state over short time intervals is then dramatically enhanced.
As a result, two atoms  near such a resonance  oscillate between
essentially uncorrelated pairs of atoms and diatomic molecular states.
Experimentally such a situation can be created, provided the atoms can exist
in two different spin states permitted by spin-orbit coupling. Different spin
alignments of the two atoms (in form of spin singlet or triplet states) give
rise to different inter-atomic interactions, which lead to either
bound or scattering states. Favoring the one over the other can be monitored
by applying a  magnetic field which shifts the respective interaction
potentials with respect to each other.

We shall illustrate  in this Letter, that similar effects are bound to play
a role in  strongly coupled electron-lattice systems with polaronic  and
bipolaronic charge carriers. The recently much studied high-$T_c$ cuprates,
the colossal magneto-resistance manganates, the nickelates as well
as the fullerenes are examples where such effects are likely to occur.  It is
however not our aim here to devote the present study to such specific
cases and we shall limit ourselves to purely theoretical questions.

It has been known for quite some time that a single electron in a
lattice which is coupled to local lattice deformations undergoes,
upon changing the electron-lattice coupling and/or the
adiabaticity ratio, a rapid crossover between a well defined
itinerant quasi-particle (describable in terms of a weak coupling
Born-Oppenheimer approach) and a predominantly incoherent
excitation in form of a self-trapped small
polaron\cite{Eagles-71,Toyozawa-71,Shore-73,deRaedt-83}. A similar
feature applies to  a single pair of electrons in such a lattice.
In the narrow crossover regime this involves two-electron states
in a superposition of uncorrelated itinerant electrons in an
essentially undeformed lattice and  localized bipolaronic bound
states, i.e., electron pairs which are self-trapped in their
strongly  deformed local lattice environments. How such effects,
involving a  single electron or a single pair of such electrons,
will be carried over into a dense electron system has occupied
much theoretical work over the past decades\cite{Varenna-05}. A
great variety of different and highly sophisticated many-body
techniques has been employed to shed light on the polaron physics
in such dense systems. Particular features which emerge from such
studies are the metal-insulator transition\cite{MIT} driven by
electron pairing into bipolarons and the transitions driven by a
stripping off of the lattice polarization upon  increasing the
density of charge carriers\cite{Hohenadler-03}.  Our goal here is
to unravel the underlying physics behind such crossover phenomena,
respectively  phase transitions,  which is to some extent hidden
in the  extremely complex numerical studies sofar used in this
context. We shall show that the phenomenon leading up to such a
crossover behavior is caused by the formation of resonating
bipolaron states, according to a mechanism bearing great
similarities with the Feshbach resonant pairing in atomic gases.

Let us illustrate this physics on the basis of a model system
composed of small plaquette clusters consisting of (i) a ring of
four atoms on which the electrons behave in a tight binding
itinerant fashion, and (ii) a central site made up of a strongly
deformable diatomic molecule described by a harmonic oscillator.
Such clusters can be imagined being linked together to form a
lattice by sharing the atoms at the corners of the square
plaquettes. This makes the system a bipartite lattice with
itinerant electrons on one sublattice and localized bipolarons on
the other one. Our system can thus be envisaged as a band of tight
binding electrons with a localized bipolaronic level inside this
band. Electrons with an energy close to this bipolaronic level
will scatter in and out of the central molecular sites and thus
acquire polaronic features, at the same time pairing up in form of
short lived bipolarons. This mechanism has an evident similarity
with the Feshbach resonant pairing in ultracold atomic gases. The
electrons moving on the sublattice made up of the rings mimic the
incident scattering states of the atoms. When the electrons  meet
on the central site, they are momentarily bound into an electron
pair (in the language of atomic physics, via so called closed
channel collisions), before fluctuating back onto the ring. In a
Feshbach low energy binary collision process the wave function of
the corresponding Schr\"odinger equation is a superposition of (i)
the incident scattering states, characterized by spin singlet
configurations of the pair of atoms with a repulsive inter-atomic
interaction, and (ii) closed channel states, characterized by spin
triplet configurations with an attractive inter-atomic
interaction. In the polaron problem the equivalent of these spin
configurations are two configurations of the oscillator on the
molecular site: (i) one with a practically undisplaced
intra-molecular distance, inciting the electrons to remain on the
ring and hence uncorrelated  and (ii) one with a displaced
shortened intra-molecular distance, inciting the electrons to pair
up on the molecular site.
\begin{figure}
\onefigure{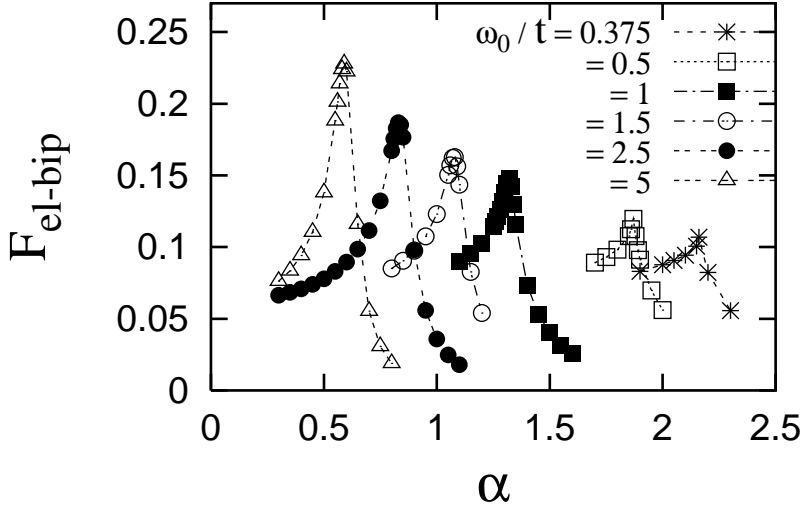} \caption{The static correlator $F_{\rm
el-bip}(\tau=0)$ as a function of $\alpha$ for various
adiabaticity ratios $\omega_0/t$. Invariably the peak position
coincides with $\alpha_c$.} \label{Fig1}
\end{figure}

The essential physics of this resonance problem  is already
contained in the electronic and vibrational properties of a single
isolated cluster. This is because the efficiency of converting the
bipolarons into resonating electron pairs involves predominantly
short range interactions and thus concerns only the immediate
vicinity of the polaronic sites, i.e., the atoms on the rings
around them. Let us now pursue our investigation on the basis of a
single cluster, for which we have the following Hamiltonian:
\begin{eqnarray}
H & = & - \, t \sum_{i \ne j = 1 ... 4, \,\sigma} \left[
c^+_{i\sigma} c_{j\sigma}
\, + \, h.c.  \right]  \nonumber \\
& & - \, t' \sum_{i=1 ... 4, \,\sigma} \left[ c^+_{i\sigma}
c_{5\sigma} \, + \, h.c.  \right] \, + \, \Delta \, \sum_{\sigma}
c^+_{5\sigma} c_{5\sigma}  \nonumber \\
& & + \, \hbar\omega_0 \left[a^+_5 a_5 + {1\over 2}\right] \, -
\hbar\omega_0\alpha \, \sum_{\sigma} \, c^+_{5\sigma}
c_{5\sigma}\, \left[a_5 + a^+_5\right] \; .
\end{eqnarray}
Here $c^{(+)}_{i\sigma}$ denote the annihilation (creation)
operators for electrons with spin $\sigma$ on the site $i$, $i=
1...4$ denoting the sites on the ring and $i=5$  the central
molecular site, while $a^{(+)}_5$ denote phonon annihilation
(creation) operators on the latter. $t$  and $t'$ denote the
hopping integrals on the ring and between the ring and the central
site, respectively. $\Delta$ is the bare energy level of the
central site, $\alpha$ the dimensionless electron-phonon coupling
constant and  $\omega_0$ the bare local phonon frequency.
Expressing the energies in units of $t$, we choose $t'=0.5$ and
$\Delta=1.5$. Approximating the full set of phonon states on the
central site by a truncated Hilbert space, we diagonalize the
above Hamiltonian for the case of two electrons. Describing the
electrons on the ring in terms of their wave vectors $q =
0,\pm\pi/2,\pi$ we find that the ground state energy $E_0(\alpha)$
is very close to $E_0(\alpha=0)$ up to a value of $\alpha$ where
the energy of the localized bipolaron state $\varepsilon_{\rm
BP}(\alpha) = 2\Delta - 4\alpha^2\hbar\omega_0$ drops below the
ground state energy $E_0(\alpha=0)$ of the itinerant electrons on
the ring. This value of $\alpha$, $\alpha_c =
\frac{1}{2}\sqrt{(2\Delta -E_0(\alpha=0) ) / \hbar \omega_0}$
determines the crossover regime between itinerant and localized
behavior. For $\alpha \geq \alpha_c$ $E_0(\alpha)$ drops off like
$\varepsilon_{\rm BP}(\alpha)$. This crossover is characterized by
a strong enhancement around $\alpha_c$ (see Fig.1) of the static
correlation function $F_{\rm el-bip}=\langle
0|c^+_{5\uparrow}c^+_{5\downarrow} c^{\phantom+}_{q=0
\downarrow}c^{\phantom +}_{q=0 \uparrow}|0\rangle$,
 which measures the efficiency to convert two uncorrelated electrons on
the ring into a bipolaron on the central site (here $|0 \rangle$
denotes the two-particle ground state).
\begin{figure}
\onefigure{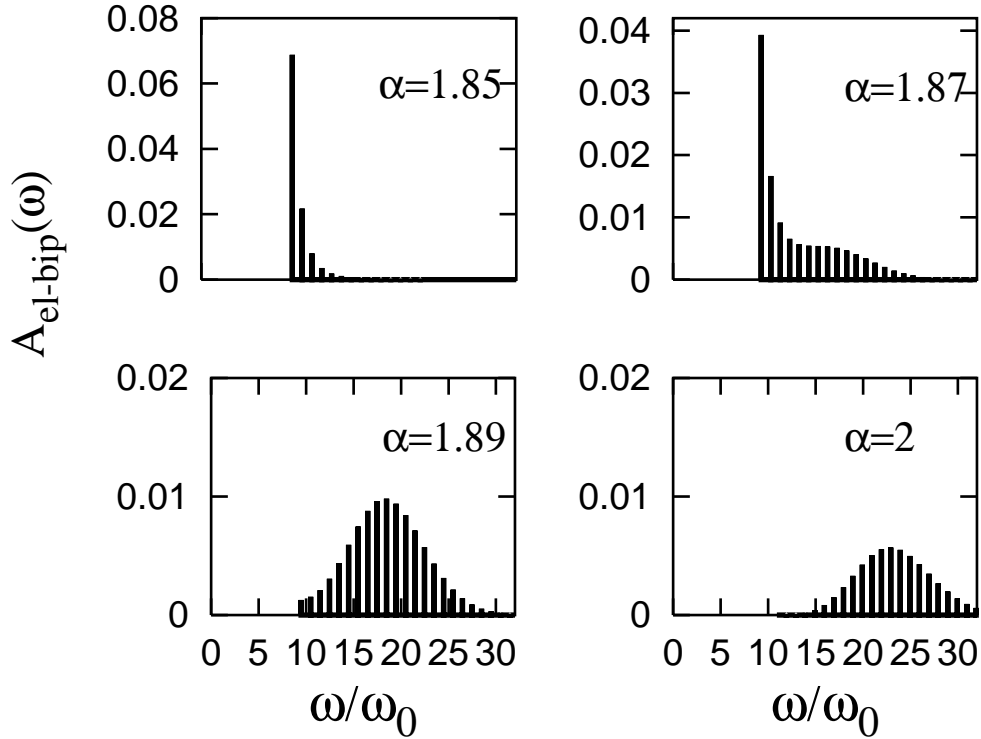} \caption{The dynamical correlator
$A_{\rm el-bip}(\omega)$ for a fixed adiabaticity ratio
$\omega_0/t=0.5$ and  various electron-phonon couplings $\alpha$.
It describes the scattering of two itinerant electrons into
localized bipolaronic states and keeps track of resonant
scattering near $\alpha_c=1.87$.} \label{Fig2}
\end{figure}
Unlike the case of the Feshbach resonance in atomic collision
processes, the interaction leading to the conversion of itinerant
electrons into localized bipolarons and vice-versa is time
dependent and involves the building up and dismantling of a local
lattice deformation. This effect is visible in the behavior of the
time dependent features of the above correlator $F_{\rm el-bip}$,
described by the corresponding spectral function
\begin{equation}
A_{\rm el-bip}(\omega) = - \frac{1}{\pi} \; {\rm Im}\,
\int_0^{\infty} d\tau e^{i\omega \tau}
\langle0|c^+_{5\uparrow}(\tau)c^+_{5\downarrow}(\tau) c^{\phantom
+}_{q=0 \downarrow}(0)c^{\phantom +}_{q=0 \uparrow}(0)|0\rangle.
\end{equation}
This is illustrated in Fig. 2 in the adiabatic limit (for a
specific choice of $\omega_0/t= 0.5$) and for several values of
the electron-phonon coupling  $\alpha$ in the crossover regime. We
notice a spectral behavior which in the weak coupling regime
($\alpha \leq 1.85$) corresponds to well defined coherent features
describing two itinerant uncorrelated electrons on the ring, being
dressed by a few harmonic overtones coming from the accompanying
phonons. In the strong coupling regime ($\alpha \geq 1.89$), we
observe a spectral behavior typical of localized bipolarons which
approaches a spectral  distribution given by
$e^{-4\alpha^2}(2\alpha)^{2n}/n!\,\delta(\omega + \varepsilon_{\rm
BP}(\alpha) - n\hbar\omega_0)$. Near and at the crossover, around
$\alpha \simeq \alpha_c = 1.87$, we notice a superposition of a
coherent contribution, coming from the two electrons freely moving
on the ring, and an incoherent one coming from those two electrons
when they stick together for a short time on the central site in
form of a bipolaron.
\begin{figure}
\onefigure{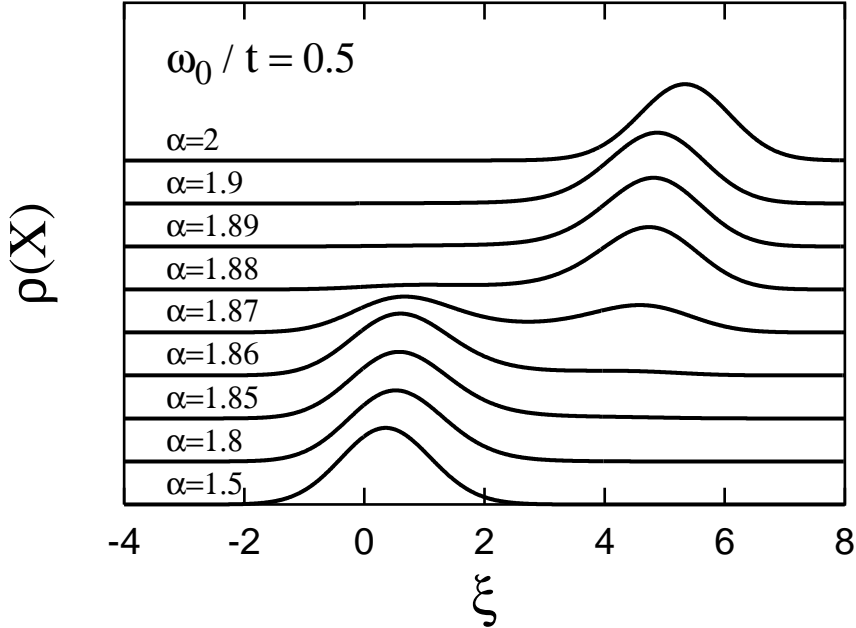} \caption{The pair distribution function
$\rho(X)$ for the oscillator site as a function of the
intra-molecular distance $\xi = X\sqrt{M\omega_0/\hbar}$ for a
fixed adiabaticity ratio $\omega_0/t=0.5$ and various
electron-phonon couplings $\alpha$. The bipolaron resonance at
$\alpha_c=1.87$ is manifest in the appearance of a double peak
structure in a narrow regime around $\alpha_c$.} \label{Fig3}
\end{figure}

Let us now further specify  the features of the crossover regime.
Going through the resonance upon  increasing $\alpha$, the
oscillator on the central site will show modifications in the
intra-molecular distance $\langle0|\widehat{X}|0\rangle$ with
$\widehat{X} = \sqrt{\hbar/2 M \omega_0}(a^+_5 +a^{\phantom +}_5)$
(M denoting the mass of the molecule) as well as in the phonon
frequency $\omega_0$. For weak coupling this distance remains
essentially unshifted from its $\alpha=0$ value, while for strong
coupling it tends to that associated with the presence of a
localized bipolaron, i.e., $X_0 = 4\alpha\sqrt{\hbar/2 M
\omega_0}$. At the resonance we observe a superposition of the
two. This behavior is illustrated in Fig. 3, where we plot the
pair distribution function (PDF) $\rho(X)  = \langle 0|\delta(x -
X)|0 \rangle$, measuring the probability to find a certain value
$X$ for the intra-molecular distance. This is done by taking the
real-space representation of the two-particle ground state
$|0\rangle$ in terms of the excited harmonic oscillator
wavefunctions. The plot in Fig.3 signals the resonance by the
appearance of a double peak structured PDF. Below and above this
resonance the PDF is singly peaked either around $X \simeq 0$ or
$X \simeq X_0$. The analogous feature in Feshbach resonant pairing
of atoms is a distribution function measuring the probability for
singlet and triplet components of the atomic pair wave function.
Above the resonance it would be given by predominantly triplet
bound pairs and below it by singlet components of the two incident
atoms. At the resonance it would correspond to a superposition of
the two.

In order to get a resonance behavior in the two-particle channel making the
{\it a priori} uncorrelated particles stick together for some substantial
time (which in the case of the atomic gases in traps can be as long as a
fraction of a second), the dynamics of the attractive interaction in the
present polaron problem must be significantly slowed down as compared to the
characteristic frequency $\omega_0$. This is precisely what happens, as
can be seen by inspection of the phonon spectral function
\begin{eqnarray}
B(\omega) = -\frac{1}{\pi}\; {\rm Im}\, \int_0^{\infty} d\tau
e^{i\omega\tau} \langle0|X(\tau)X(0)|0\rangle
\end{eqnarray}
whose behavior is illustrated in Fig. 4.
\begin{figure}
\onefigure{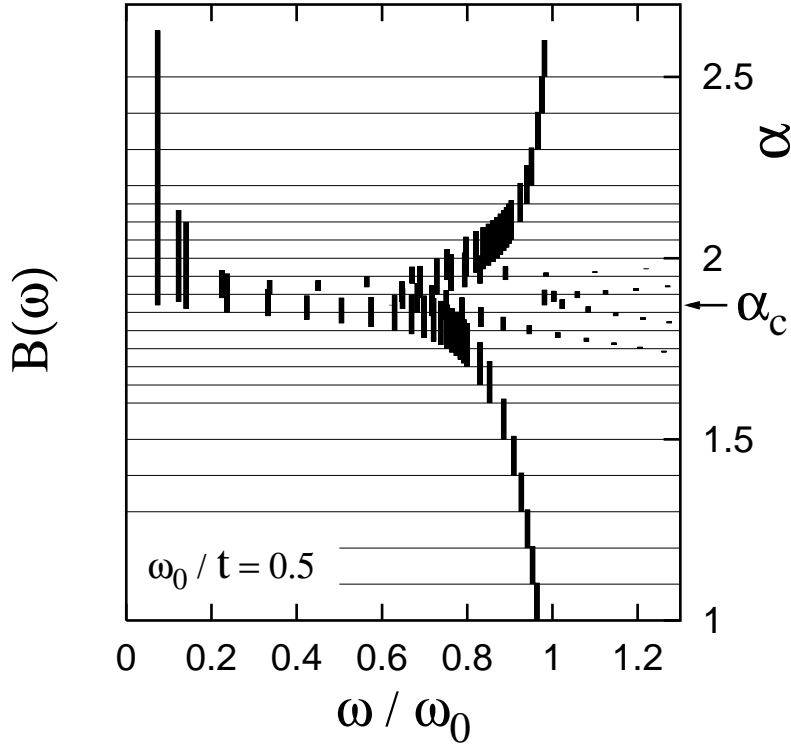} \caption{The phonon spectral function
$B(\omega)$ for a fixed adiabaticity ratio $\omega/t= 0.5$ and
various electron-phonon couplings $\alpha$, as indicated on the
right $y$-axis. Upon approaching the resonance at $\alpha_c$, it
shows a substantial softening and a simultaneous increase in
spectral weight (given by the length of the vertical bars).}
\label{Fig4}
\end{figure}
The bare phonon frequency $\omega_0$ is reduced to $\omega^*_0$ by
more than an order of magnitude  upon approaching the resonance at
$\alpha = \alpha_c$, while at the same time its spectral weight is
significantly enhanced. These are the features which control the
physics leading to bipolaron resonant states. The lifetime of such
resonating bipolarons is of the order of $\pi/\omega^*_0$, which
is about an order of magnitude longer than the characteristic time
of the intrinsic molecular oscillation, i.e., $\pi/\omega_0$. The
electrons then oscillate  with a frequency $\omega^*_0$ between an
uncorrelated behavior on the ring and a trapped bipolaronic one on
the molecular site. This fluctuation of the electrons in and out
of the polaronic site is associated with the oscillation of the
intra-molecular distance, a sluggish mode of the oscillator
switching back and forth between two different bond lengths of the
molecule. Increasing $\alpha$ slightly above $\alpha_c$ keeps the
electrons bound as bipolarons, while a decrease of $\alpha$ below
$\alpha_c$ makes them turn into itinerant quasi-particles. The
mechanism of tuning the system via $\alpha$ between the different
regimes is fully equivalent to the tuning across the Feshbach
resonance in atomic gases by applying an external magnetic field.
Experimentally, tuning in electron-phonon coupled systems could be
achieved by either changing the electron density or the phonon
frequency $\omega_0$ (via isotope substitution of the molecular
units responsible for bipolaron formation) such as to modify the
difference between the chemical potential and the energy level of
the bipolarons.

This Letter aimed to attract attention to a possible resonant
pairing in polaronic systems in the intermediate coupling regime.
A particularly suited scenario for that is represented by systems
composed of a lattice of metal atoms being surrounded by their
ligand environments. The latter provide the local dynamical
deformations which lead to selftrapped polaronic entities. Our
contention is that global features (such as superconductivity or
insulating behavior) emerge from the local physics already
contained in appropriately chosen clusters. An infinite system
composed of such clusters, interconnected by electron hopping, can
be treated by real space contractor plaquette renormalization
group methods, which have been successfully applied to the Hubbard
problem\cite{Auerbach-02} in order to capture the relevant local
physics of the correlation problem for an infinite lattice system.

The hallmark of such resonant pairing is a strongly enhanced
conversion between localized bipolarons on the ligand and
itinerant electrons in their immediate vicinity - a feature which
we illustrated here on the basis of an isolated cluster. For an
infinite lattice, composed of such interconnected clusters, the
electrons which partially fill a tight binding band will overlap
with the level of the localized bipolarons, describing the
situation when the electrons find themselves momentarily on the
ligands. Hence there will always be a degeneracy of uncorrelated
pairs of itinerant electrons and localized bipolarons such that
such a resonance behavior can occur. For this reason the
particular choice of our parameter $\Delta$, fixing the level of
the localized electrons on the ligands will not influence the
resonant pairing phenomenon on a qualitative level since by
changing the electron lattice coupling (or alternatively by
changing the band filling) we can always bring the energy level of
the localized bipolarons $\varepsilon_{\rm BP}$ inside the filled
part of the tight-binding electron band.

Finally, we should point out that the phonon softening, indicative
of the resonant pairing, is qualitatively similar to what has been
obtained for the single-impurity Anderson-Holstein model analyzed
by NRG\cite{Hewson-02a} as well as for the Holstein model at half
filling within DMFT\cite{Koller-04,Capone-05}. Given the
electron-phonon interaction $-\hbar \omega_0 \alpha \sum_{\sigma}
(c^+_{5\sigma} c_{5\sigma} - \frac{1}{2}) [a_{5}+a^+_{5}]$ used in
those studies and an impurity level coinciding with the Fermi
energy, the resulting degeneracy of empty and doubly occupied
impurity sites favors a Kondo resonance in charge space, whose
dynamics shows up in phonon softening similar to that of the
Feshbach-like resonance discussed here. Away from this limiting
situation this Kondo resonance disappears\cite{Hewson-02a}, while
the resonance discussed in this Letter is expected to remain away
from half filling, for a bare energy $\Delta$ of the polaronic
sites not coinciding with the Fermi level, and for the usual form
of the electron-phonon interaction.


\end{document}